\pdfoutput=1
\documentclass{JINST}

\title{Measurement of electrical properties of electrode materials for the bakelite Resistive Plate Chambers}

\author{K.~K.~Meghna$^{a,b}$, A.~Banerjee $^{c}$, S.~Biswas$^{ c,d}$\thanks{Corresponding
author.}, S.~Bhattacharya$^b$, S.~Bose$^b$, S.~Chattopadhyay$^c$, G.~Das$^c$, C.~Marick$^b$, S.~Saha$^b$ and Y.~P.~Viyogi$^c$\\
\llap{$^a$}Institute of Mathematical Sciences,\\
  Chennai-600 113, India\\
\llap{$^b$}Saha Institute of Nuclear Physics,\\
  1/AF Bidhan Nagar, Kolkata-700 064, India\\
\llap{$^c$}Variable Energy Cyclotron Centre,\\
  1/AF Bidhan Nagar, Kolkata-700 064, India\\
\llap{$^d$}GSI Helmholtzzentrum f\"ur Schwerionenforschung GmbH,\\
  Planckstrasse 1, D-64291 Darmstadt, Germany\\
  E-mail: \email{saikat.ino@gmail.com}}

\abstract{Single gap (gas gap 2 mm) bakelite Resistive Plate Chamber (RPC) modules of various sizes from 10 cm $\times$ 10 cm to 1 m $\times$ 1 m have been fabricated, characterized and optimized for efficiency and time resolution. Thin layers of different grades of silicone compound are applied to the inner electrode surfaces to make them smooth and also to reduce the surface resistivity. In the silicone coated RPCs an efficiency $>$ 90\% and time resolution $\sim$ 2 ns (FWHM) have been obtained for both the streamer and the avalanche mode of operation. Before fabrication of detectors the electrical properties such as bulk resistivity and surface resistivity of the electrode materials are measured carefully. Effectiveness of different silicone coating in modifying the surface resistivity was evaluated by an instrument developed for monitoring the I-V curve of a high resistive surface. The results indicate definite correlation of the detector efficiency for the atmospheric muons and the RPC noise rates with the surface resistivity and its variation with the applied bias voltage. It was also found that the surface resistivity varies for different grades of silicone material applied as coating, and the results are found to be consistent with the detector efficiency and noise rate measurements done with these RPCs.}


\begin{document}

\section{Introduction}
The Resistive Plate Chambers (RPCs), first developed by Santonico et al. \cite{RSRC81} using bakelite are used extensively in high energy physics and neutrino physics experiments. The RPCs are being considered for the following reasons a) relatively low cost of materials used in making RPCs, b) robust fabrication procedure and handling and c) excellent time and position resolution. Primarily used for generating fast trigger for muon detection \cite{GB94}, time of flight (TOF) \cite{RCRS88,AB03} measurement, and tracking capabilities in multi layer configurations, they are successfully used in BELLE \cite{AA02}, BaBar \cite{BABAR95}, BESIII \cite{TBD13}, and several LHC experiments (ATLAS,CMS etc.) \cite{ATLAS,CMS}. RPCs are used in neutrino experiments like OPERA where its excellent time resolution and tracking capabilities are exploited \cite{OPERA}. 

In the proposed India-based Neutrino Observatory (INO), the RPCs have been chosen as the prime active detector for the detection of muons (produced through the interaction of neutrinos) in an Iron
Calorimeter (ICAL) \cite{INO06}.  INO is being planned to determine the neutrino oscillation parameters precisely in the 3-flavor mixing scenario using atmospheric neutrinos. For effective separation of up-coming and down-going neutrinos and background rejection, ICAL requires highly efficient and sensitive detectors with $\lesssim$~2~ns time resolution. During the last few years significant work on the prototype silicone coated bakelite based RPC has been carried out at SINP/VECC for INO-ICAL. Bakelite RPC detectors of various sizes from 10 cm $\times$ 10 cm to 1 m $\times$ 1 m have been fabricated, characterized and optimized for efficiency and time resolution, and are reported earlier \cite{SBose109,SB109,SB209,SB309,SB110,SBTh}. A particular grade of bakelite (P-120, NEMA LI-1989 Grade XXX) has been used to build as many detectors. For RPC electrodes, careful choice of materials, smoothness of surfaces to avoid localization of excess charges, surface treatment to reduce the surface resistivity or providing alternate leakage path for post-streamer recovery are adopted in the major high energy physics experiments. The current paper deals with the development of instruments and methods for the measurement of bulk resistivity of  the materials and surface resistivity of the inner electrode surfaces of the bakelite RPCs. Incidentally, these developments are complimentary to the development of glass-based RPCs in other collaborating institutes of the proposed INO project.
 \section{Measurement of bulk resistivity}

\begin{figure}[htb!]
\begin{center}
\includegraphics[scale=0.63]{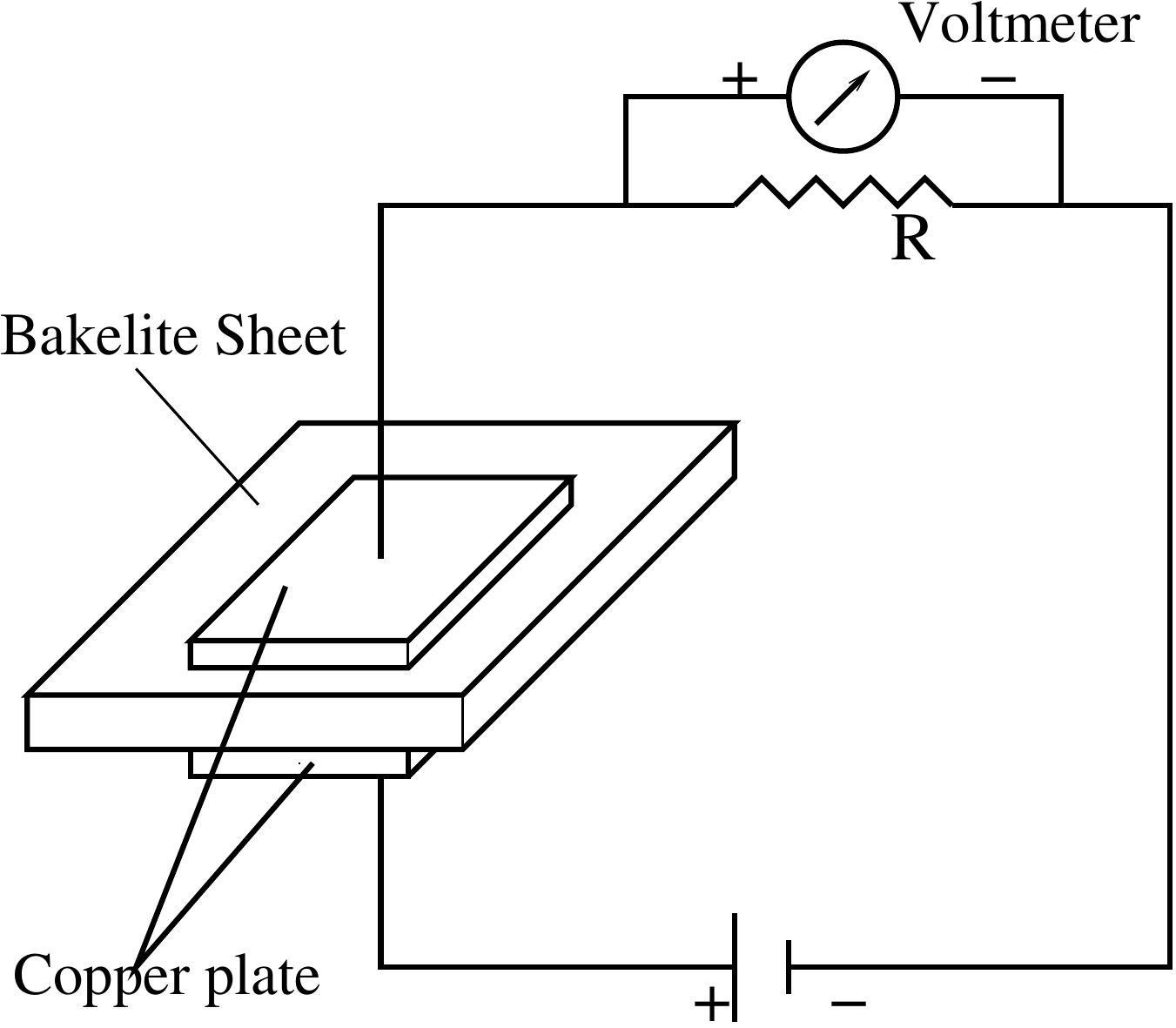}
\caption{\label{bulkres}Schematic diagram of the bulk resistivity
measurement setup.}\label{bulkres}
\end{center}
\end{figure}

The bakelite sheets are phenolic resin bonded paper laminates. The bulk resistivity (volume resistivity) of the electrode plates of the RPC is an
important parameter \cite{GA04,GB93}. The high resistivity helps in
controlling the time resolution, counting rate and also prevents the
discharge from spreading through the entire gas volume \cite{RSRC81,RC93,HC98}.
We have measured the bulk resistivities of the bakelite
sheets via the measurement of leakage current. Figure~\ref{bulkres} shows the schematic diagram of the set up for the bulk resistivity measurement. Bakelite samples are cut into 3~cm~$\times$~3~cm sizes. The sample was sandwiched between two copper plates of dimension 2~cm~$\times$~2~cm and thickness of 1.5~mm. The copper sheets are pressed on to the bakelite plates using two 10~cm~$\times$~10~cm glass epoxy pieces with a hole at the centre and screwed at the 4 corners (for tightening and not shown in the picture). Two cables have been soldered on to the copper plates and connected to the LeCroy 1458 high voltage power supply and a 33 M$\Omega$ resistance (R) in series. By varying the applied voltage from the power supply, measured voltages across R are used to obtain the leakage current and the bulk resistivity of the bakelite sheet.

\begin{figure}[htb!]
\begin{center}
\includegraphics[scale=0.5]{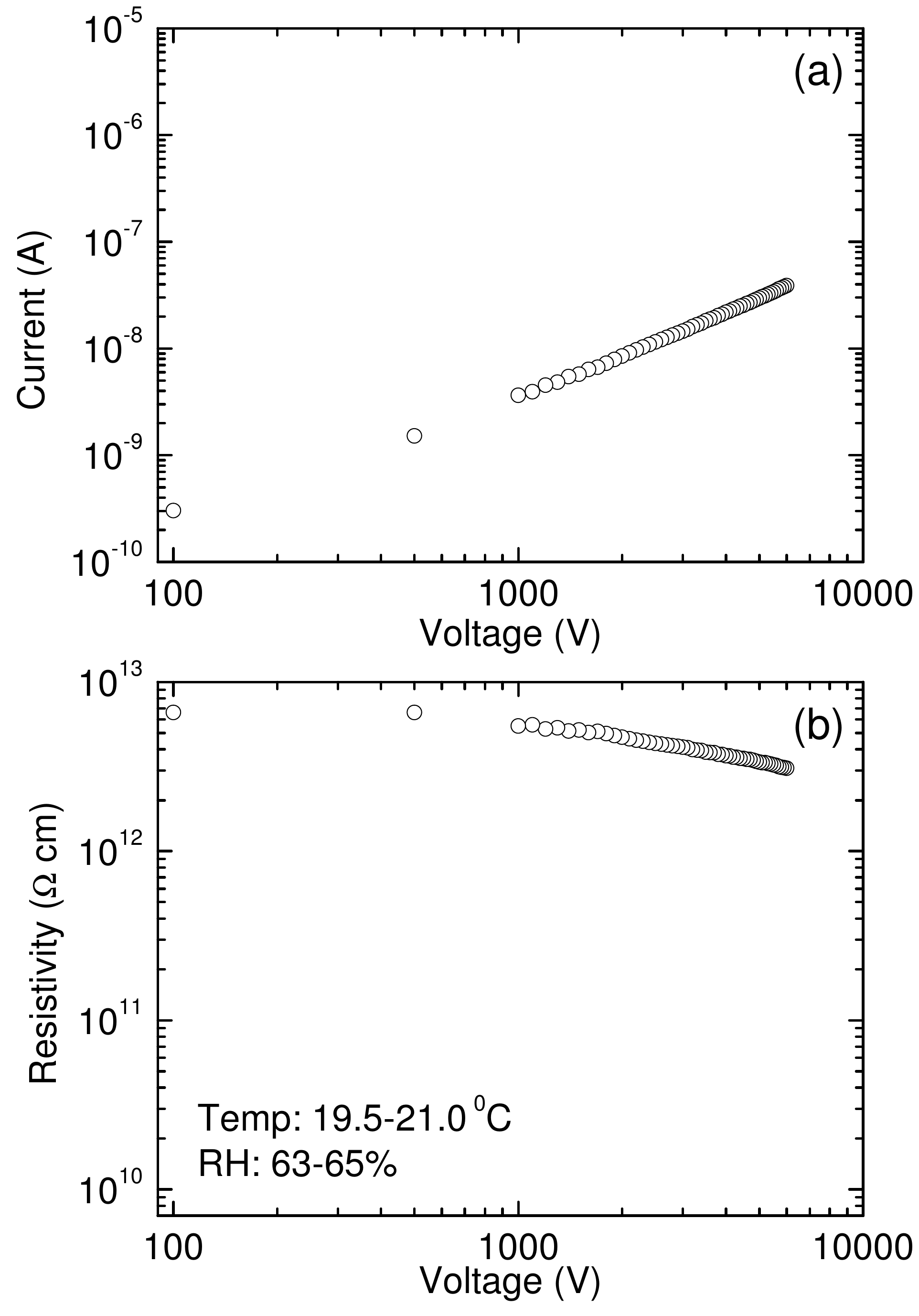}
\caption{\label{fig1}(a) The current $\&$ (b) the volume resistivity~($\rho$) as a function of the applied voltage for P-120 grade bakelite.}
\end{center}
\end{figure}

This measurement has been performed in a temperature and humidity controlled room where
the RPCs have also been tested, since the resistivity of highly hygroscopic bakelite depends on humidity \cite{LB12}. These two parameters have been monitored during the measurement and were kept almost constant during the entire period of experiment. Measured temperature and relative humidity during the experiment were 19.5-21.0 $^{\circ}$C and 63-65\% respectively. The bulk resistivity of the P-120 grade material at 4 kV is 3.67 $\times$ 10$^{12}$ $\Omega$~cm with an accuracy of 0.8\%. The variation of leakage current and volume resistivity~($\rho$) with applied voltage are shown in Figure~\ref{fig1}~(a) and (b) respectively for P-120 grade bakelite sheet.
\section{Measurement of surface resistivity}


\begin{figure}[htb!]
\begin{center}
\includegraphics[scale=0.9]{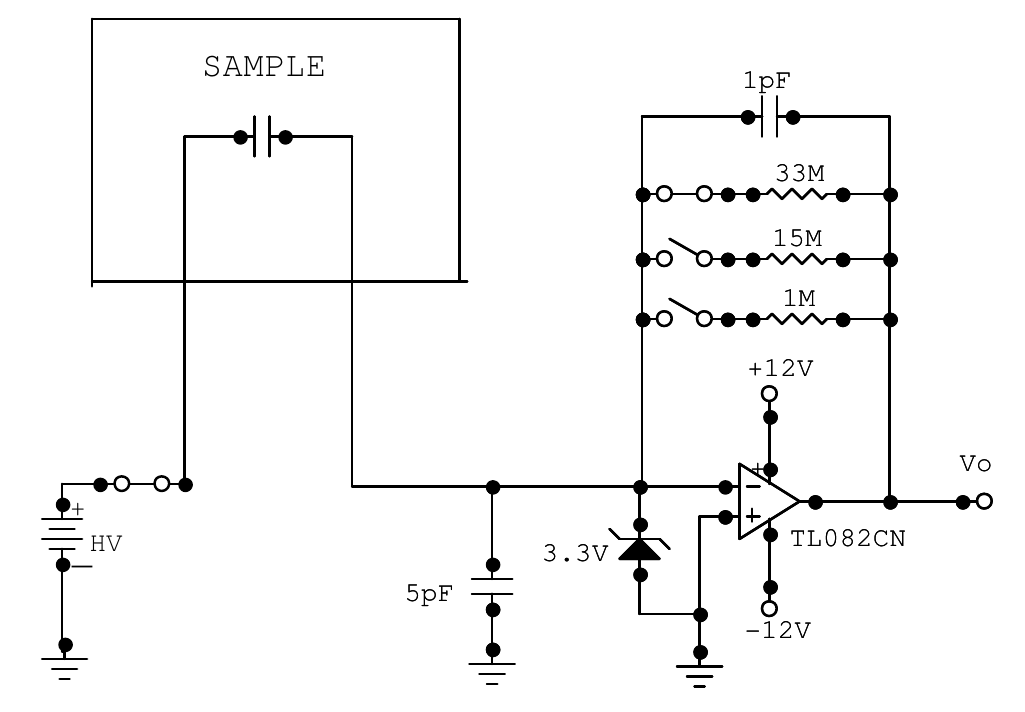}
\caption{\label{surckt}Block diagram of the surface
resistivity measurement setup.} \label{surckt}
\vspace{0.8 cm}
\includegraphics[scale=0.4]{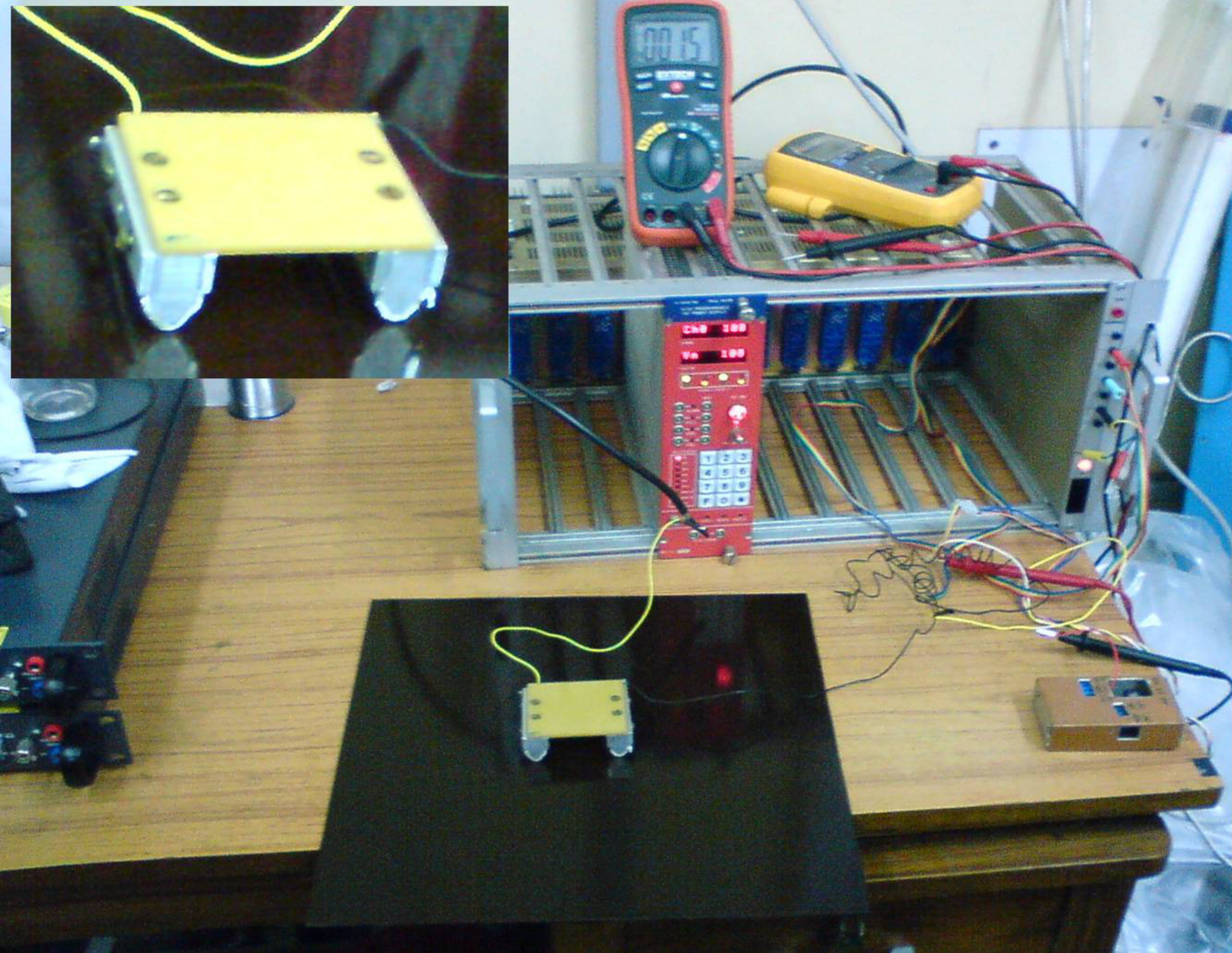}
\caption{\label{surface_res_setup}Experimental setup for surface resistivity measurement. The inset shows  the aluminium jig used for the purpose.} \label{surface_res_setup}
\end{center}
\end{figure}


Thin layers of different grades of silicone compounds are applied to the inner electrode surfaces to make them smooth and also to reduce the surface resistivity. Before fabrication of each RPC module, the surface resistivity of bakelite sheets and also of silicone coated surfaces are measured using the set-up shown schematically in the  Figure~\ref{surckt}. The actual experimental set-up is shown in the Figure~\ref{surface_res_setup}. The set-up consists of a jig with two aluminium bars having V-shaped sections (shown in the inset to Figure~\ref{surface_res_setup}) and soft-padded conducting edges at the bottom, which are placed on the surface under measurement. The bars, forming the opposite sides of a square shape, were mounted on a G-10 insulating plate having very high resistivity ($>$ 10$^{14}$ $\Omega$/$\Box$). The length of the aluminium bars and their separation were same (5 cm). A current to voltage converter circuit, made out of TL082CN FET input OPAMP, with provisions to cover 3 decades of surface resistivity measurement ($\sim$ 10$^{10}$ - 10$^{12}$ $\Omega$/$\Box$), was made.

Measurements were done on the inner surfaces of the bakelite electrodes (silicone coated or uncoated) before assembly of the RPCs. A DC bias voltage $\sim$ 50 - 600 volt was applied on the jig, and the leakage current ($\sim$ nA/pA) flowing across the terminals of the jig through the bakelite surface was measured. The surface resistivity was obtained from the leakage current and the applied bias voltage.

The bakelite sheets used for making the RPC electrodes were obtained in two batches, having different surface textures. The samples from the bakelite sheets were scanned under an Atomic Force Microscope (AFM) to determine their roughness. The AFM plots for the two samples (rough (a) and smooth (b)) are shown in the Figure~\ref{afm} side by side for comparison. The average roughness for the 'rough' and the 'smooth' surfaces were found to be 100 nm and 18 nm respectively. The surface resistivity was measured for these two surface grades, along with the coatings applied on them.

\begin{figure}[htb!]
\begin{center}
\includegraphics[scale=0.35]{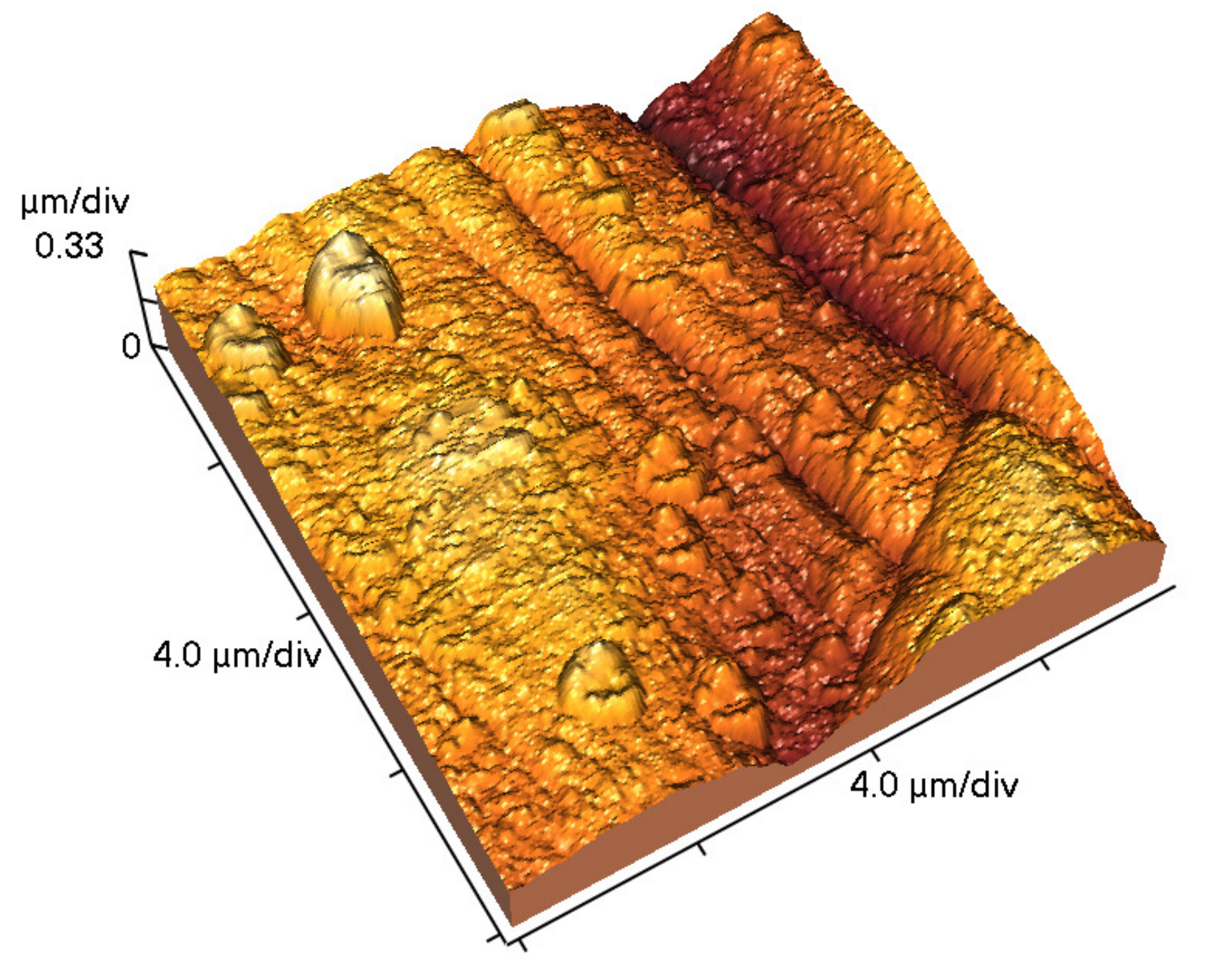}
\includegraphics[scale=0.35]{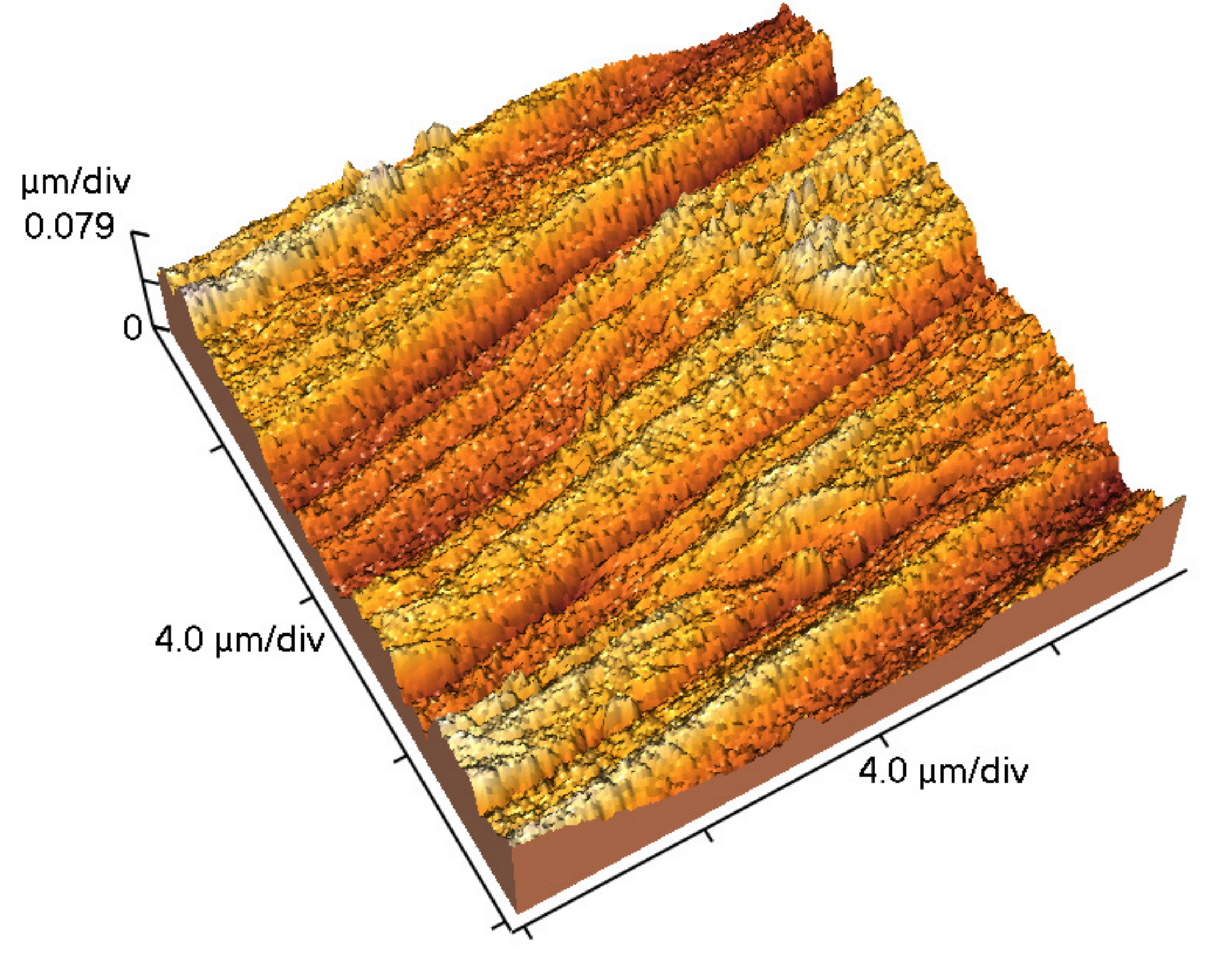}
\hskip -1.0cm
\includegraphics[scale=0.30]{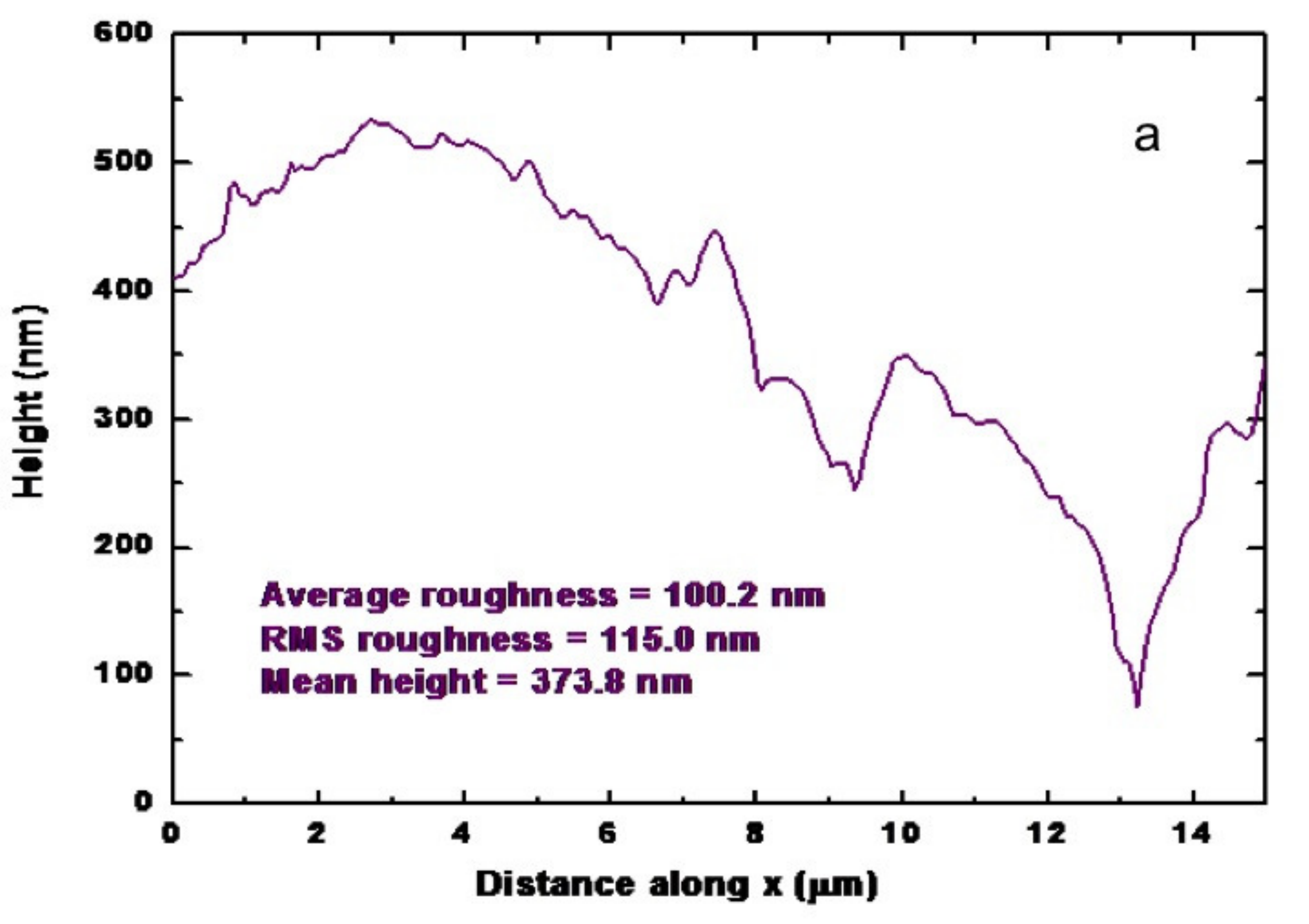}
\hskip -0.2cm
\includegraphics[scale=0.29]{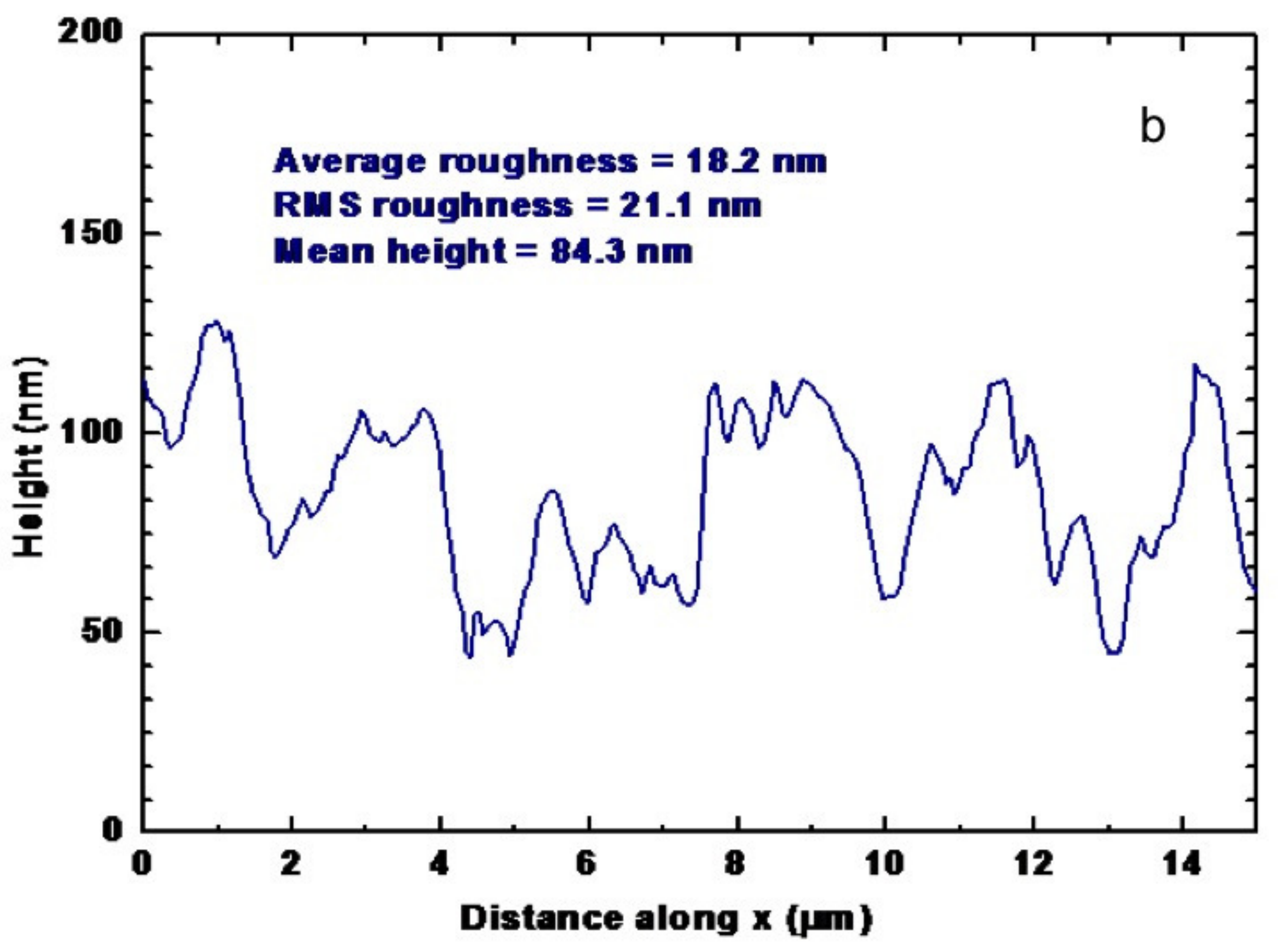}
\caption{\label{afm}AFM image and topography of AFM data for the (a) rough  and (b) smooth bakelite surface morphological structure.}\label{afm}
\end{center}
\end{figure}

\begin{figure}[b!]
\begin{center}
\includegraphics[scale=0.59]{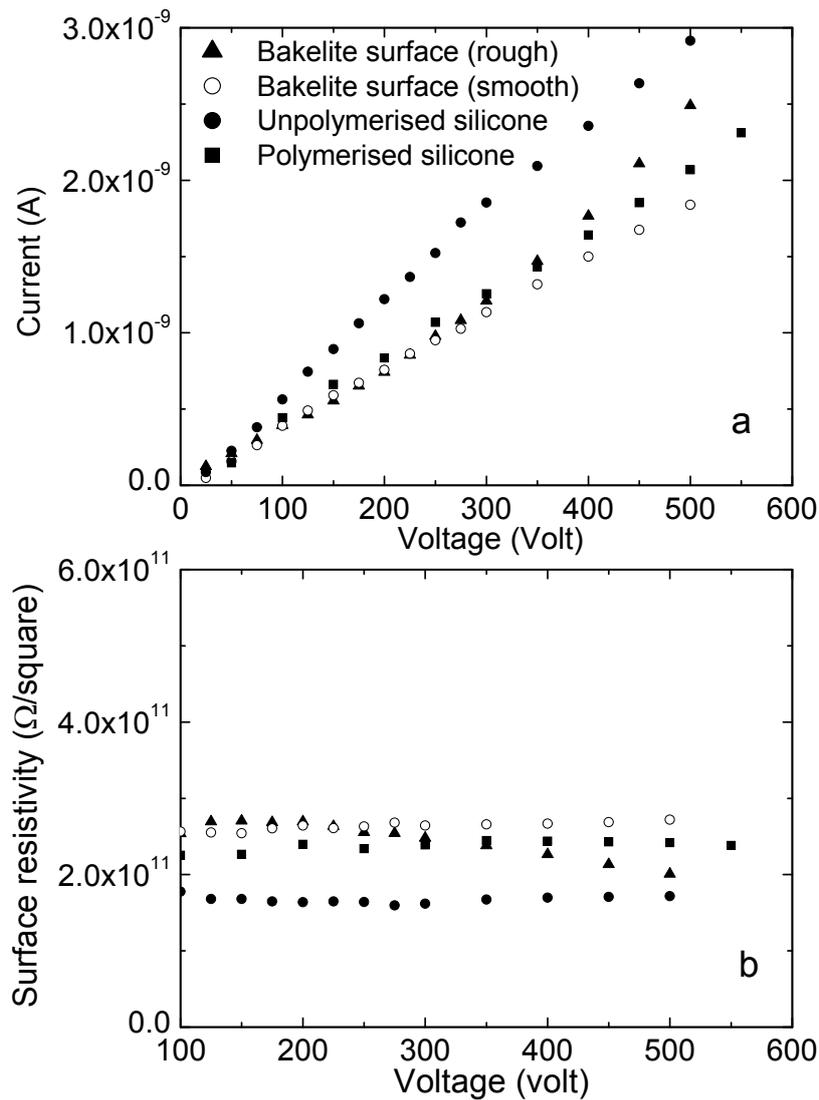}
\caption{\label{surcurrent}(a) The surface current $\&$ (b) the surface resistivity versus the applied
voltage for different samples.}\label{surcurrent}
\end{center}
\end{figure}
The variation of surface current and that of surface resistivity with the applied voltage for different coated and uncoated bakelite electrodes are shown in Figure~\ref{surcurrent}~(a) and \ref{surcurrent}~(b) respectively. A look at the I-V plots for the surface resistivity of uncoated bakelites having two different surface textures (rough and smooth) shows a non-linear trend for the rough surface, resulting in an apparent reduction of surface resistivity at higher bias voltage. This is likely to be correlated with the relatively high occurrence of micro-discharge across the surface due to the roughness \cite{NM12}. This finding correlates with the reduction of efficiency and increase of RPC noise rate at higher bias voltages for the bakelite RPCs, made with the rough variety in our earlier studies \cite{SB109}. In addition, comparison of the two different grades of silicone coating (unpolymerized and polymerized) on the bakelite electrode surface shows that 1) the surface resistivity is less by a factor of $\sim$ 2 for the coated surface compared to the uncoated surface, and 2) the surface resistivity for the polymerized silicone coating is $\sim$ 1.5 times higher than that for the unpolymerized variety. Lower value of surface resistivity is expected to help in reducing the space charge effect because of quicker dissipation of accumulated charge through the surface layer.

It may be noted that RPCs made by silicone coated (unpolymerised) bakelite plates as electrodes have been tested earlier for a long duration showing $\sim$~96\% efficiency for a period of operation of more than 130 days \cite{SB109}. The time resolution for the RPCs was found to be $\sim$~2~ns \cite{SB309} and the measured average charge content per pulse is $\sim$~100~pC at an applied high voltage of 8~kV in the streamer mode of operation \cite{SB110}.

\section{Conclusions and outlook}
The bulk resistivity of the electrode material of the RPC is an important parameter. The measured value of the bulk resistivity is useful for the proper choice of the electrode material for RPC. This measurement is very essential before fabrication of RPC modules. The bulk resistivity measurement set-up, described here can measure high resistivity $\sim$~10$^{10}$ - 10$^{12}$ $\Omega$ cm. The bulk resistivity of the P-120 grade material is measured to be 3.67 $\times$ 10$^{12}$ $\Omega$~cm at 4 kV and at 20 $^{\circ}$C with an accuracy of 0.8\%.

An instrument is developed to measure the surface resistivity of high resistive ($\sim$~10$^{10}$ - 10$^{12}$~$\Omega$/$\Box$) surface such as glass and bakelite. Using this instrument, measurements were done on the inner surfaces of several bakelite electrodes (silicone coated or uncoated) before fabricating the detectors. The surface resistivity of the silicone coated (unpolymerised) surface was found to be less by a factor of 2 compared to the uncoated surface and by a factor of $\sim$~1.5 compared to the polymerised silicone coated variety. These findings correlate well with our earlier observation on the long term stability of performance of the unpolymerized silicone coated bakelite RPC detectors.


\begin{thebibliography}{9}

\bibitem{RSRC81}R. Santonico, R. Cardarelli, \emph{Development of resistive plate counters}, {\emph{Nucl. Inst. and Meth.} {\bf187}, (1981) 377}.
\bibitem{GB94}Gy. L. Bencze et al., \emph{Study of resistive plate chambers for muon detection at hadron colliders}, {\emph{Nucl. Inst. and Meth. A} {\bf340} (1994) 466}.
\bibitem{RCRS88}R. Cardarelli et al., \emph{Progress in resistive plate counters}, {\emph{Nucl. Inst. and Meth. A} {\bf263} (1988) 20}.
\bibitem{AB03}A. Blanco et al., \emph{Resistive plate chambers for time-of-flight measurements}, {\emph{Nucl. Inst. and Meth. A} {\bf513} (2003) 8}.
\bibitem{AA02}A. Abashian et al., \emph{The Belle detector}, {\emph{Nucl. Inst. and Meth. A} {\bf479} (2002) 117}.
\bibitem{BABAR95}BaBar Technical Design Report, BaBar Collaboration,
SLAC Report SLAC-R-95-457, March 1995.
\bibitem{TBD13}The BESIII Detector, IHEP-BEPCII-SB-13, IHEP,
Beijing.
\bibitem{ATLAS}ATLAS Technical Design Report, Muon Spectrometer,
CERN/LHCC/97-22, Geneva, 1997.
\bibitem{CMS}CMS - Technical Proposal, CERN/LHCC/94-38, December 1994.
\bibitem{OPERA}M. Guler et al., \emph{OPERA, an appearance experiment to
search for $\nu_{\mu}$-$\nu_{\tau}$ oscillations in the CNGS beam},
CERN/SPSC 2000-028.
\bibitem{INO06}INO Project Report, INO/2006/01, June 2006, $\langle$http://www.imsc.res.in/$\sim$ino/$\rangle$.
\bibitem{SBose109}S. Bose et al., \emph{Control system for a four-component gas mixing unit}, {\emph{Nucl. Instr. and Meth. A} {\bf602} (2009) 839}.
\bibitem{SB109}S. Biswas et al., \emph{Performances of linseed oil-free bakelite RPC prototypes with cosmic ray muons}, {\emph{Nucl. Instr. and Meth. A} {\bf602} (2009) 749}.
\bibitem{SB209}S. Biswas et al., \emph{Development of linseed oil-free bakelite resistive plate chambers}, {\emph{Nucl. Instr. and Meth. A} {\bf604} (2009) 310}.
\bibitem{SB309}S. Biswas et al., \emph{Study of timing properties of single gap high-resistive bakelite RPC}, {\emph{Nucl. Instr. and Meth. A} {\bf617} (2010) 138}.
\bibitem{SB110}S. Biswas et al., \emph{Performances of silicone coated high resistive bakeliteRPC}, {\emph{Nucl. Instr. and Meth. A} {\bf661} (2012) S94}.
\bibitem{SBTh} Saikat Biswas, PhD Thesis, \emph{Development of high resolution gas filled detector for high energy physics experiments}, 2011, University of Calcutta.
\bibitem{GA04}G. Aielli et al., \emph{Electrical conduction properties of phenolic-melaminic laminates}, {\emph{Nucl. Instr. and Meth. A} {\bf533} (2004) 86}.
\bibitem{GB93}G. Bencivenni et al., \emph{A glass spark counter for high rate environments}, {\emph{Nuclr. Inst. and Meth. A} {\bf332} (1993) 368}.
\bibitem{RC93}R. Cardarelli A. Di Ciaccio and R. Santonico, \emph{Performance of a resistive plate chamber operating with pure CF$_3$Br}, {\emph{Nucl. Instr. and Meth. A} {\bf333} (1993) 399}.
\bibitem{HC98}H. Czyrkowski et al., \emph{New developments on resistive plate chambers for high rate operation}, {\emph{Nucl. Instr. and Meth. A} {\bf419} (1998) 490}.
\bibitem{LB12}L. Benussi et al., \emph{Study of gas purifiers for the CMS RPC detector}, {\emph{Nucl. Instr. and Meth. A} {\bf661} (2012) S241}.
\bibitem{NM12} N. Majumdar et al., \emph{Study on Surface Asperities in Bakelite-RPC},
\pos{PoS(RPC2012)026}.

\end{thebibliography}
\end{document}